\documentstyle[12pt,aasms]{article}
\tighten
\slugcomment{Submitted to the Astrophysical Journal}
\begin{document}
\def \m{\ifmmode M_\odot\else M$_\odot$\fi}
\def \etal {{\it et~al.}}
\def\ctr{\centerline}
\def \lta {\mathrel{\vcenter
     {\hbox{$<$}\nointerlineskip\hbox{$\sim$}}}} % less than approx
\def \gta {\mathrel{\vcenter
     {\hbox{$>$}\nointerlineskip\hbox{$\sim$}}}} % greater than approx

\title{OCCLUSION EFFECTS AND THE DISTRIBUTION OF
INTERSTELLAR CLOUD SIZES AND MASSES}
\author{John Scalo and A. Lazarian}
\affil{Astronomy Department, University of Texas, Austin,
    TX 78712-1083}
\begin{abstract}
The frequency distributions of sizes of ``clouds" and ``clumps" within
clouds are significantly flatter for extinction surveys than for CO spectral
line surveys, even for comparable size ranges.  A possible explanation is the
blocking of extinction clouds by larger foreground clouds (occlusion), which
should not affect spectral line surveys much because clouds are
resolved in velocity space along a given line of sight.  We present a
simple derivation of the relation between the true and occluded size
distributions, assuming clouds are uniformly distributed in space or the
distance to a cloud comples is much greater than the size of the complex.
Because the occlusion is dominated by the  largest clouds, we find that
occlusion does not affect the measured size distribution except for
sizes comparable to the largest size, implying that occlusion is not
responsible for the discrepancy if the range in sizes of the samples is large.
However, we find that the range in sizes for many of the published
observed samples is actually quite small, which suggests that
occlusion does affect the extinction sample and/or that the discrepancy could
arise from the different operational definitions and selection effects involved
in the two samples.  Size and mass spectra from an IRAS survey (Wood \etal\
1994) suggest that selection effects play a major role in all the
surveys.  We  conclude that a reliable determination of the ``true" size and
mass spectra of clouds will  require spectral line surveys with very high
signal-to-noise and sufficient resolution and sampling to cover a larger range
of linear sizes, as well as careful attention to selection effects.
\end{abstract}


\section{Introduction}

        The mass spectrum of density fluctuations, defined in various
operational ways as ``clouds", is an important function that must
be related
to the processes by which clouds form and evolve and to the mass
spectrum
of stars that form within these clouds.  A fairly large and growing
number
of studies of (mostly) molecular clouds yield a differential mass
spectrum
which, if fit by a power law, has a form $f(m)\sim m^{-\gamma}$,
with
$\gamma \sim -1.5\pm 0.2.$  These studies are primarily based
on masses derived from
column densities inferred from $^{13}$CO spectral line
observations and linear
size.  Some of the results are for surveys that cover a significant
area of
the Galactic disk, e.g. the second quadrant survey of Casoli \etal\
(1984),
who find $\gamma=-1.4$ to $ -1.6$ in both the Perseus and
Orion arms, the $^{12}$CO first
quadrant survey by Sanders \etal\ (1985) who find $\gamma
\sim -1.6$ using virial
masses, and the recent comparison of 204 inner- and outer-
galaxy molecular
clouds by Brand and Wouterloot (1995), who find $\gamma=-1.6$
for outer Galaxy
clouds and $\gamma=-1.8$ for all 204 clouds.  These surveys
together cover a mass
range from about 100 \m\ to over $10^{6}$ \m, although each
individual study
generally covers a much smaller mass range over which a power
law is an adequate fit.  Other work
has concentrated on the mass spectrum of ``clumps" within
individual clouds complexes, and find
similar mass spectra in regions as different in star formation
properties as the
Maddalena-Thaddeus cloud (Williams, deGeus and Blitz 1994),
which shows no evidence for star
formation, the $\rho$-Oph core region (Loren 1989 as revised by
Blitz 1993), which is forming
low- to intermediate- mass stars, the Rosette Molecular Cloud
(Williams \etal\ 1994, Williams and
Blitz 1995), the M17SW cloud (Stutzki and Gusten 1990), and the Orion region
(Lada,
 Bally, and Stark 1991, Tatematsu \etal\ 1993), all of which are
actively forming stars up to
large masses, and  even
lower-mass clumps in MBM 12, a molecular cloud that is not
gravitationally
bound (Pound 1994).  All these studies give $\gamma\sim
-1.5\pm 0.3$
(the flattest being
the Williams and Blitz result for the Rosette cloud, with
$\gamma\sim-1.3$).

        However there is a probable discrepancy when these results
are
compared with studies of the mass spectra of clouds derived using
extinction surveys, which are also based on masses from sizes and
column
densities.  If the distribution of sizes is given by $f(r)\sim
r^{\alpha}$ and
the cloud internal density $n$ is related to size by $n\propto
r^{p}$, then the mass
spectrum is $f(m)\sim m^{\gamma}$ with $\gamma = (\alpha-p-
2)/(3+p)$.  Estimates of p are
uncertain and vary from (at least) $p\sim-1.2 $(see Scalo 1985,
1987) to $p=0$ (sizes or
masses uncorrelated with density, e.g. Casoli \etal\ 1984,
Williams and
Blitz 1995), or that the correlation is at least in part an artifact
due to
selection effects (Scalo 1990). The spectral line studies mentioned
earlier
give values of $\alpha$ around -2 to -2.5, based either on the
published size data when
available or on the above transformation between mass and size
spectra.

        Scalo (1985) presented the frequency distribution of angular
surface areas of dark clouds from the catalogues of Lynds (1962)
and
Khavtassi (1960).  The resulting size spectrum, if fit by a power
law, has
$\alpha\sim -1.4\pm0.2$, much flatter than the size
distributions inferred from the
spectral line surveys.  The implied mass spectra ($\gamma \sim-
1.2\pm0.3$) seem
significantly flatter than the spectral line mass spectrum, but,
because of
the above relationship between size and mass spectra (which gives
$\Delta\alpha=(2-3)\Delta\gamma$ for $p=-1$ to 0), and because
size is a directly measured
quantity in both types of studies, the discrepancy is more clearly
seen in the size spectrum.  Feitzinger and St\"{u}we (1986) studied the
statistics
of the combined sample of Lynds clouds and their own Southern
dark cloud survey, and found a distribution of areas proportional to
(area)$^{-1}$.  The
corresponding size spectrum has $\alpha=-1$.  This gives a mass spectrum index
of -1 for any p.
Thus the discrepancy with the molecular line survey size or mass spectra is
even larger.
 Other
published studies of mass spectra based on extinction are  not so clear, but
point in the same
direction, especially for lower-mass clumps (e.g. Bhatt
\etal\  1984) for Lynds
clouds in Orion, $\rho$ Oph, and Taurus.  Drapatz and
Zinnecker (1981) give size and mass
spectra for several samples based on both extinction and CO.

        In the present paper we examine the possibility that this
flatter
size spectrum seen in extinction is due to the effects of occlusion
(smaller clouds being hidden behind large clouds) on the
extinction
studies; this effect would not affect the spectral line studies nearly
as
much because in that case two clouds along the same line of sight
can be
distinguished in velocity space.  (Of course occlusion in velocity
space
can also occur; we discuss this briefly in \S\ 3 below.) We derive
an expression for the real
size distribution of clouds in terms of the measured distribution
that is affected by binary
occlusion and derive the range of parameters over which the
difference in size spectra between
the two approaches can be reconciled.

A relation between the distribution of physical sizes of clouds and
their
angular sizes is established in \S\ 2, while a relation between
the actual distribution of angular sizes and the distribution
measured in the presence of
occlusion is presented  in \S\ 3.

\section{``Apparent" sizes of clouds}

Consider that $N_{1}(l)$ is the `real' size distribution and
$N_{2}(\theta)$ is the
``apparent" angular size distribution, without accounting for
occlusion.
In this seciton we define the relation between $N_{1}(l)$ and
$N_{2}(\theta)$. This
problem is similar to the one discussed in Feitzinger and St\"{u}we (1986).
 Due to the geometry of diverging lines of sight,
clouds
with the same physical size but at different distances from the
observer
will fall into different ranges of apparent angular sizes.  As a first
approximation, assume the ``true" properties of clouds to be
independent of
the distance from the observer. This is probably reasonable for
observations in the
galactic plane and of nearby individual cloud complexes (e.g.
Taurus, Oph, Chameleon, Orion,. .
.)

In our model, the distribution of clouds at distance $r$
is given by the product $\varrho(D)N_{1}(l)$, where
$\varrho(D)$ is
the total density of clouds at distance D, and we take $N_1(\ell)$
normalized to unity. Then
the number of clouds within  the distance interval $D$, $D+dD$
is $\varrho(D)\omega
D^{2}{\rm d}D$, where $\omega$ is the solid angle.
  Within this volume, the clouds with sizes from
$D\theta$ to
$(\theta+{\rm d}\theta)D$, where $\theta$ is the angular size of
clouds,
will contribute to the apparent angular cloud
distribution $N_{2}(\theta)$. The total number of ``projections"
with
angular sizes ($\theta$, $\theta+{\rm d}\theta$) within the solid
angle $\omega$
can be found by integrating
$\varrho(D)\omega D^{2}N_{1}(D\theta)D{\rm d}\theta{\rm d}D$
over the line of sight. Therefore,
\begin{equation}
N_{2}(\theta)\omega {\rm d}\theta=\omega\int_{D_{\rm
min}}^{D_{\rm max}}
D^{3}\varrho(D)N_{1}(D\theta){\rm d}D{\rm d}\theta
\end{equation}
A change of variables $D\theta=x$ results in
\begin{eqnarray}
N_{2}( \theta )=-\frac{1}{ \theta^{4}}\int_{\theta D_{\rm min}}^{
\theta D_{\rm max}}
x^{3}\varrho\left(\frac{x}{ \theta }\right)N_{1}(x){\rm d}x
\label{*}
\end{eqnarray}
Assuming $\varrho(x)=$ constant, differentiation gives
\begin{eqnarray}
\frac{1}{D_{\rm max}}\left(N_{2}( \theta ) \theta^{4}\right)'&=&
 \theta^{3}D_{\rm max}^{3}\varrho N_{1}( \theta D_{\rm
max})\nonumber\\
&-& \theta^{3}D^{3}_{\rm min}\varrho N_{1}( \theta D_{\rm min})
\label{4}
\end{eqnarray}

For power law $N_{1}(\ell)=N_{1}( \ell_{\rm min})
\left(\frac{ \ell }{ \ell_{\rm min}}\right)^{-\gamma}$,
the first term is the most important if  $\gamma<3$, whereas if
 $\gamma>3$, the second term dominates. The cases of greatest
interest here have $\gamma<3$.
Whenever the
second term is negligible and  $N_{1}(\ell)$  is a power-law
distribution,
$N_{2}(\theta)$ is also power-law with the same index.

Similarly for $D_{\rm min}=0$,
\begin{equation}
N_{1}( \theta D_{\rm max})=\frac{1}{ \theta^{3}D_{\rm
max}^{4}\varrho}
(N_{2}( \theta ) \theta^{4})'
\end{equation}
 and for any other $D_{\rm min}$, the power-law distribution
$N_{1}(l)$
entails a  power-law distribution $N_{2}(\theta)$ with equal slope.
Therefore  the index of the size distribution is not affected by
the differing distances of the
clouds in the sample, and the index of the angular size
distribution is the same as the index of
the linear size distribution.  An exception occurs for a delta
function linear size
distribution, i.e. when all clouds have the same size.  In that case
the apparent angular size
distribution varies as $\theta^{-4}$ (see Bhatt \etal\ 1984).  In
what follows, we therefore
identify $N_2(\theta)$ with the ``real'' distribution of sizes, with
the understanding that
clustering of clouds and gradients in the number density of clouds
with distance could alter this
identification. Obviously, if the distance to a cloud complex is much greater
than the extention of the complex, statistics of the ``real'' size
distribution and the ``apparent''angular size distribution coinside.

\section{Occlusion effect}

$N_{2}(\theta)$ is the projected apparent angular size distribution
of clouds when occlusion is
ignored; i.e. it is the angular size distribution corresponding to the
``real" linear size
distribution. If occlusion is ``switched on," some of smaller clouds
are hidden  behind (or in
front) of bigger ones. Let
$N_{3}(\theta)$  be the distribution of projections in the presence
of occlusion. Then
\begin{equation}
\pi N_{3}(\theta )\theta^{2}\omega{\rm d}\theta
\end{equation}
is the angular area covered by cloud projections with sizes within
the
range $\theta$, $\theta +{\rm d}\theta$. The part of the sky not
covered by
cloud projections with angular sizes greater than $\theta $ is
\begin{equation}
1-\frac{\pi}{A}\int_{\theta }^{\theta_u}N_{3}(x)x^{2}{\rm d}x
\end{equation}
where A is the angular area covered by the survey  and $\theta_u$
is the upper size limit for
the sample. Therefore the number of cloud projections of angular
size $\theta$ that are {\it not}
occluded by larger  clouds is
\begin{equation}
N_{2}(\theta)\omega{\rm d}\theta
\left(1-\frac{\pi}{A}\int_{\theta }^{\theta_u}N_{3}(x)x^{2}{\rm
d}x\right)
\end{equation}
Since this is the number of clouds that is seen, equating this to
$N_3(\theta)\omega d\theta$
gives
\begin{equation}
N_{3}(\theta )=N_{2}(\theta )
\left(1-\frac{\pi}{A}\int_{\theta }^{\theta_u}N_{3}(x)x^{2}{\rm
d}x\right)
\end{equation}
The real size distribution $N_{2}(\theta)$ can therefore be derived
from the apparent (occluded
size distribution $N_3(\theta)$ from
\begin{equation}
N_2(\theta)=\frac{N_3(\theta)}{1-
\frac{\pi}{A}\int^{\theta_u}_\theta N_3(x)x^2dx}
\end{equation}
The second term in the denominator is just the fraction of the
survey area A covered by clouds
with sizes greater that $\theta$.  The largest value this fraction
can have occurs at
$\theta=\theta_\ell$, the minimum size detected in the survey,
for which the second term is the
total area filling factor of clouds detected in the survey ($<1$).

It is also possible to derive the observed distribution $N_3(\theta)$
that would result from a
given real distribution $N_2(\theta)$, as shown in the Appendix.
However, that formulation is
not as useful for the purposes of the present paper because the
solution involves the unknown
properties of the real distribution.

To illustrate the properties of the $N_2-N_3$ relation,
assume that the observed occluded distribution is a power law,
$N_3(\theta)=c_3\theta^{-\gamma_3}$.  Then
\begin{equation}
N_2(\theta)=\frac{c_3\theta^{-\gamma_3}}{1-
\frac{\pi}{A}\frac{c_3}{(3-\gamma_3)}\left(\theta_u
^{-r_3+3}-\theta^{-\gamma_3+3}\right)}
\end{equation}
The total areal filling fraction is
\begin{equation}
f_{3,tot}=\frac{1}{A}\int^{\theta_u}_{\theta_\ell}
N_3(\theta)\pi\theta^2d\theta=\frac{\pi
c_3}{A(3-\gamma_3)}\left(\theta_u^{-\gamma_3+3}-\theta_\ell^{-
\gamma_3+3}\right).
\end{equation}
The second term is negligible for $\theta_\ell\ll\theta_u$ and
$\gamma_3<3$.  So, from eqs. (10)
and (11), we see that for $\theta$ significantly smaller than
$\theta_u,\
N_2(\theta)=N_3(\theta)/(1-f_{3,tot})$; i.e. for small clouds the real
number of clouds is larger
than the observed number by a factor ($1-f_{3,tot})$, but the
power law index is unaffected.
The probability of a small cloud to be hidden by a large cloud is
independent of its size if its
size is much smaller than $\theta_u$ because the areal filling is
dominated by the largest
clouds (if $\gamma_3<3)$.

To see this more clearly, consider the local logarithmic slope of the
real distribution at size
$\theta$ (i.e. the exponent of a local power law fit at that size).
 From eq. (10) we obtain
\begin{eqnarray}
\gamma_2(\theta)=\frac{d\ell nN_2(\theta)}{d\ell n\theta } &=&
\gamma_3+\frac{\pi
c_3}{A}\frac{\theta^{-\gamma_3+3}}{\left[1-\frac{\pi c_3}{A(3-
\gamma_3)}\left(\theta_u^
{-\gamma_3+3}-\theta^{-\gamma_3+3}\right)\right]}
\nonumber\\
&\equiv& \gamma_3+\Delta\gamma(\theta)
\end{eqnarray}
The maximum value of the change in exponent
$\Delta\gamma(\theta)$ occurs for $\theta$ near
$\theta_u$, at which size $\Delta\gamma(\theta)=\pi
c_3\theta_u^{-\gamma_3+3}/A\approx(3-\gamma_3)f_{3,tot}$ (for
$\theta_\ell\ll\theta_u$ and
$\gamma_3<3$).  If $f_{3,tot}\approx0.5$, as is typical for dark
cloud surveys (not selected
according to opacity class or size), then the dark cloud power law
$\gamma_3\sim1.4$ gives
$\Delta\gamma\approx1.6f_{3,tot}\sim0.8$.  While this is about
the value needed to reconcile the
extinction size distribution with the spectral line size distribution,
it only occurs very close
to $\theta_u$.  At smaller $\theta$, say $x\theta_u(x<1),\
\Delta\gamma$ is reduced by a factor
of $x^{-\gamma_3+3}\sim x^{1.6}$ for the parameters chosen.  So
even for clouds half or a third
of the size of the largest clouds, $\Delta\gamma$ is too small to
account for the discrepancy, and
for $x=0.1$, $\Delta\gamma$ is essentially negligible.

The same considerations hold even if the observed distribution
$N_3(\theta)$ is not a power law,
as long as it is not locally too steep ($\gamma>3$):  the real size
distribution tracks the
observed distribution (although at larger amplitude) except for
sizes close to $\theta_u$, at
which sizes the real distribution is steeper than the observed
distribution.

We would be tempted to conclude that occlusion cannot account
for the discrepancy, except for
the fact that the range in sizes in the observed surveys is actually
quite small.  For both
types of surveys, the cloud masses are proportional to the square
of some characteristic size
times a column density, so the range in sizes, which is a directly
observed datum, only
corresponds to the square root of a given range in the masses
(which is what is usually
displayed).  Since, for the published spectral line surveys of clumps
within cloud complexes,
power laws are only good fits over a limited mass range (limited by
small numbers at the largest
masses and resolution incompleteness and other effects at small
masses), usually a factor of
10--100, the range in sizes is not very large.  The range in sizes for
a few early surveys is
listed in Drapatz, and Zinnecker (1984).  The range in {\it sizes} for
the line surveys of Stutzki and Gusten (1990), Lada \etal\ (1991), Tatematsu
\etal\
 (1993), Williams
\etal\ (1984) and Williams  and Blitz (1995) is less than a
factor of 10, although the range in mass
used to derive the mass spectra is larger in some of the
surveys.  This suggests that the mass distributions derived from spectral
line surveys will be very sensitive to the
definition of, and systematic uncertainties in the measurement of,
cloud sizes.  For the
extinction sample the range of sizes over which the power law size
spectrum is applicable is less
than a factor of about 10 in all cases, even for the full sample of the
Lynds and Khavtassi surveys, and
various selection effects come into play at smaller and larger
masses (see Scalo 1985, \S\
III.B.2. for a discussion).

Thus we conclude that the discrepancy between size distributions
derived from extinction surveys
and spectral line surveys {\it may} be due to occlusion effects in
the extinction surveys
because the minimum size is not much smaller than the
maximum size in both types of surveys,
{\it or} because of different operational definitions of size in the
two types of surveys.
Actually these two possibilities are not independent because the
size range is related to how
clouds are defined.  It is worth pointing out that in some of the
spectral line surveys the
noise level is so large that the surveys are really only observing the
``tips of the mountain
range" if the column density map is thought of as a 2-dimensional
surface with height equal
to column density.  For example, in the Rosette data (Blitz and
Stark 1986), the rms noise is only
about a factor of 2 smaller than the average {\it peak} line temperature, so
the cloud sizes may be
severely affected.  For the dark clouds, identification of the cloud
boundary is usually much
less affected by ``noise" (in this case fluctuation in star densities),
except for the
lowest-opacity clouds.  Thus even though the line surveys have
the advantage of separating
clouds in velocity space it is not clear that they give more realistic
size and mass spectra
compared to extinction surveys.

Evidence that the empirical cloud mass spectra are sensitive to selection
effects comes from the following two examples.

Clemens and Barvainis (1988) compiled a catalogue of isolated small dark cloud
(``globules")
identified on POSS plates and compiled properties based on their CO
observations.  For clouds
with mean size larger than 3.5 arcmin (smaller size clouds are probably
affected by
incompleteness), we can fit the frequency distribution of angular sizes, and
hence linear sizes
if the clouds are uniformly distributed in distance,
by $f(r)\sim r^{-2}$, which gives a power law mass
spectrum with $\gamma=-1.5$ to -1.7 for $p=-1$ or 0.  These clouds were
selected to be small and
isolated, so occlusion should not be important.  Since this result agrees with
the molecular line
surveys, it suggests that the flatter size and mass spectra derived from
general extinction
surveys are products of occlusion effects,
 {\it if} selection effects are unimportant in the estimation of
properties from CO.

However the survey of 255 IRAS cloud cores by Wood, Myers, and Daugherty
(1994), which derives
sizes and masses based on IRAS 100 $\mu$m optical depth for clouds with
$A_V\gta4$ mag, yields a
frequency distribution of areas $f(A)\sim A^{-0.54}$, or $f(r)\sim r^{-0.08}$,
which is extremely
flat compared to not only the molecular line surveys, but even extinction
surveys.  For
constant column density, as Wood \etal\ find, $p=-1$, so $f(m)\sim m^{-0.54}$,
consistent with
their directly determined (from individual areas and column densities)
$f(m)\sim m^{-0.49}$.
The sizes and masses for the fits have ranges of well over 1000.  Since all the
cores are
optically thin at 100 $\mu$m, occlusion cannot be a factor; a small core behind
a larger core
would be seen as a column density enhancement of about a factor of two, because
all the cores
in the sample apparently
have about the same column density.  This result suggests that all the surveys,
whether based on extinction, molecular line, or IRAS, are affected by selection
effects.

\section{Velocity occlusion}

The same argument used above for purely spatial occlusion can be
somewhat extended to include
the effects of blending in velocity space for spectral line surveys.  In
this illustrative
example we assume that each identified ``cloud" or ``clump" (for
convenience we use the latter
term in what follows) has an internal velocity dispersion $\Delta
v(\theta)$ which is strictly
correlated with the size of the clump, as found in several surveys,
at least for clumps in which
self-gravity is important.  In that case the fraction of the total
survey volume of the data
cube AV (A = area of the survey in the plane of the sky, V = radial
velocity extent of the
survey) occupied by occluded clumps of size $\theta$ is
\begin{displaymath}
f_v(\theta)=\frac{1}{AV}\int^{\theta_u}_\theta
\pi\theta^2N_v(\theta)\Delta v(\theta)d\theta
\end{displaymath}
where $N_v(\theta)$ is the size distribution found in the (blended)
survey.  The real size
distribution is then
\begin{displaymath}
N_2(\theta)=\frac{N_v(\theta)}{1-f_v(\theta)}
\end{displaymath}
The maximum value of $f_v(\theta)$ occurs at $\theta_\ell$ and
is the total volume filling
factor of observed clumps in the data cube.  Since this number is
small for the surveys we are
aware of (see Fig. 7 in Williams and Blitz 1995), the effect of this
type of occlusion (due to
finite internal velocity dispersion of the clumps) on the derived size
distribution must be
negligible, at least for velocity resolutions much smaller than the
minimum $\Delta v$. However, this analysis does not account for the fact that
clumps with
similar {\it centroid} velocities may lie along the same line of sight.  Taking
this effect
which probably dominates the blending in velocity space, into account would
involve calculting
the probability that, for a prescribed centroid velocity distribution, two
clouds along a given
line of sight have a centroid velocity difference smaller than the sum of the
linewidths of the
two clouds (which is a function of $\theta$), a calculation which we postpone
to a later
publication.

\section{Conclusions}

%Our study shows, that occlusion affects the extinction surveys.
%The expected change in the frequency distributions of sizes of
%clouds for extinction surveys as compared to CO spectral line surveys is
%small, if the range in sizes of the sample is large. At the same
%time this change may be substantial for the published surveys
%with the very limited range of sizes and therefore our results
%indicate that the occlusion can be important.
Our study has examined the effect of occlusion on extinction surveys.  The
predicted change in
the shape of the frequency distribution of cloud sizes for extinction surveys
compared to
spectral line surveys is small, except very near the maximum cloud size.  Thus
the discrepancy
between the empirical results for the two types of surveys probably cannot be
attributed to
occlusion in the extinction survey, if the size range of both types of survey
is large.  Howver
an examination of the literature shows that many of the observed surveys employ
a very limited
range of sizes.  In these cases the discrepancy might still be due to
occlusion.  On the other
hand, some of the spectral line surveys do include clouds with a fairly large
range of sizes
(e.g. Brand and Wouterloot 1995), and these surveys do find size and mass
spectra much steeper
than the dark cloud results.  Furthermore, the IRAS cloud-core survey of Wood
\etal\ (1995)
gives size and mass spectra which are much flatter than {\it both} the
extinciton and line
survey results.  This suggests that the inferred shapes of the size and mass
spectra of clouds
are affected by the manner in which clouds are defined and by the selection and
noise effects
inherent in both types of surveys.

 \acknowledgements
This work was supported by NASA grant NAG52773.
\clearpage

\appendix
\vfill
\eject
\ctr{\bf APPENDIX}

\medskip
Rather than solve for the real distribution function in terms of the
observed (occluded)
distribution) it is possible to derive the observed distribution
$N_3(\theta)$ that would result
from a given real distribution $N_2(\theta)$.  Differentiating eq. 9
with respect to $\theta$
gives
%\setcounter{equation}{0}
%\addtocounter{equation}{A}
\begin{equation}
N_3^\prime(\theta)=N_3(\theta)\left[\frac{\pi\theta^2}{A}+\frac{N
_2^\prime(\theta)}{N_2^2(\theta)}\right]N_2(\theta)
\end{equation}
Integrating this equation, with a lower integration limit
$\theta_\ell$, gives
\begin{equation}
\frac{N_3(\theta)}{N_3(\theta_\ell)}=\frac{N_2(\theta)}{N_2(\theta_
\ell)}\
exp\left\{\frac{\pi}{A}
\int^\theta_{\theta_\ell}N_2(x)x^2dx\right\}
\end{equation}
We can obtain $N_3(\theta_\ell)/N_2(\theta_\ell)$ by imposing
the condition that the largest
cloud in the sample cannot suffer any occlusion, i.e. by
substituting
$N_3(\theta_u)=N_2(\theta_u)$ at $\theta=\theta_u$ in eq. A2.
This condition results in
\begin{equation}
\frac{N_3(\theta_\ell)}{N_2(\theta_\ell)}= exp\left\{\frac{\pi}{A}
\int^{\theta_u}_{\theta_\ell}N_2(x)x^2dx\right\}=\ exp(-
A_{tot}/A),
\end{equation}
where $A_{tot}$ is now the total area covered by all clouds in the
unoccluded (real)
distribution, and may be greater than the survey area A.  Dividing
the integral from
$\theta_\ell$ to $\theta_u$ into parts from $\theta_\ell$ to
$\theta$ and from $\theta$ to
$\theta_u$ and substituting into eq. A2 gives
\begin{equation}
N_3(\theta)=N_2(\theta)\ exp\left\{-
\frac{\pi}{A}\int^{\theta_u}_\theta N_2(x)x^2dx\right\}
=N_2(\theta)\ exp[-A(>\theta)/A],
\end{equation}
where $A(>\theta)$ is the area covered by clouds with sizes greater
than $\theta$ in the
unoccluded distribution and may xxx by greater than $A$.  For a
power law
$N_2(\theta)=c_2\theta^{-\gamma_2}$ we find
\begin{equation}
N_3(\theta)=N_2(\theta)\ exp\left\{\frac{\pi c_2}{(3-
\gamma_2)A}\
(\theta_u^{-\gamma_2+3}-\theta^{-\gamma_2+3})\right\}.
\end{equation}
The local logarithmic slope of the predicted occluded distribution
is then (assuming
$\theta_\ell\ll\theta_u$ and $\gamma_2<3)$
\begin{equation}
\gamma_3(\theta)=\frac{d\ell nN_3(\theta)}{d\ell
n\theta}=\gamma_2-
\frac{A_{tot}}{A}\left(\frac{\theta}{\theta_u}\right)^{-
\gamma_2+3}.
\end{equation}
Once again we see that although the change in local logarithm
slope may be large near
$\theta_u$, the effect becomes increasingly negligible for
$\theta\ll\theta_u$.

However this formulation is not as useful as that given in the
main text (which expressed
$N_2(\theta)$ in terms of $N_3(\theta)$) because the total
covering fraction of the real
distribution is unknown, although it can be evaluated for a model
which specifies the total
number of clouds in the distribution (again unknown from
observations).

\end{document}